\documentclass[prl,twocolumn,superscriptaddress,amsmath,amssymb]{revtex4}

\usepackage{graphicx}
\usepackage{bm}
\usepackage{multirow}

\newcommand {\Eq}[1] {Eq.~\ref{#1}}

\newcommand {\Fig}[1] {Figure~\ref{#1}}

\newcommand{\beq}{\begin{equation}}
\newcommand{\eeq}{\end{equation}}

\newcommand{\natc}{N@C$_{60}$}

\newcommand{\erscn}{ErSc$_2$N}
\newcommand{\erscnC}{ErSc$_2$N@C$_{80}$}
\newcommand{\erx}{Er$_x$Sc$_{3-x}$N}

\newcommand{\ceighty}{C$_{80}$}

\newcommand{\cstwo}{CS$_2$}

\newcommand{\beqa}{\begin{eqnarray}}
\newcommand{\eeqa}{\end{eqnarray}}

\newcommand{\ttwo}{$T_2$}

\newcommand{\tone}{$T_1$}

\newcommand{\sitea}{\emph{Conf~1}}
\newcommand{\siteb}{\emph{Conf~2}}
\begin{document}

\title{Switchable ErSc$_2$N rotor within a C$_{80}$ fullerene cage: \\ An EPR and photoluminescence excitation study
}

\author{John~J.~L.~Morton}
\email{john.morton@materials.ox.ac.uk} \affiliation{Department of Materials, Oxford University, Oxford OX1 3PH, United Kingdom}
\affiliation{Clarendon Laboratory,
Department of Physics, Oxford University, Oxford OX1 3PU, United Kingdom}

\author{Archana Tiwari}
\affiliation{Department of Materials, Oxford University, Oxford OX1 3PH, United Kingdom}

\author{Geraldine Dantelle}
\affiliation{Department of Materials, Oxford University, Oxford OX1 3PH, United Kingdom}

\author{Kyriakos~Porfyrakis}
\affiliation{Department of Materials, Oxford University, Oxford OX1 3PH, United Kingdom}

\author{Arzhang~Ardavan}
\affiliation{Clarendon Laboratory, Department of Physics, Oxford University, Oxford OX1 3PU, United Kingdom}

\author{G.~Andrew~D.~Briggs}
\affiliation{Department of Materials, Oxford University, Oxford OX1 3PH, United Kingdom}

\date{\today}

\begin{abstract}
Systems exhibiting both spin and orbital degrees of freedom, of which Er$^{3+}$ is one, can offer mechanisms for manipulating and measuring spin states via optical excitations. 
Motivated by the possibility of observing photoluminescence and electron paramagnetic resonance from the same species located within a fullerene molecule, we initiated an EPR study of Er$^{3+}$ in ErSc$_2$N@C$_{80}$. Two orientations of the ErSc$_2$N rotor within the C$_{80}$ fullerene are observed in EPR, consistent with earlier studies using photoluminescence excitation (PLE) spectroscopy. For some crystal field orientations, electron spin relaxation is driven by an Orbach process via the first excited electronic state of the $^{4}I_{15/2}$ multiplet. We observe a change in the relative populations of the two ErSc$_2$N configurations upon the application of 532~nm illuminations, and are thus able to switch the majority cage symmetry. This photoisomerisation, observable by both EPR and PLE, is metastable, lasting many hours at 20~K.
\end{abstract}

\pacs{76.30.-v, 81.05.Tp}

\maketitle


Photoswitchable structural phenomena are essential to many natural processes. In the laboratory, they have been observed in a wide range of organic~\cite{choi06} and inorganic~\cite{deboer67} species, with switching times as short as a picosecond~\cite{nagale97}. Natural photoisomerisation (such as bleaching of rod and cone retinal pigments in sight~\cite{wald67}) has been exploited to engineer optical switches which can be used to store information~\cite{liu90} and stimulate neurons~\cite{szobota07}. Fullerenes have been shown to provide remarkable nanovoids in which atoms, ions and molecules can exhibit nearly-free behaviour within a solid state environment.  Examples include the quantisied rotational states in C$_2$Sc$_2$@C$_{84}$~\cite{krause04} and the exceptionally narrow electron paramagnetic resonance (EPR) lines of atomic nitrogen in \natc~\cite{Murphy1996}. It is common for such ``endohedral'' fullerenes to possess several stable structural isomers, though no evidence of photoisomerisation in such systems has thus far been reported.

Although neither the fullerene \ceighty~nor the trimetallic nitride species \erx~($x$=0--3) is known to exist in isolation, the incarceration of such species within the fullerene cage results in the stable structure \erx@\ceighty,~where the four-atom planar unit sits in the center of the cage~\cite{stevenson04}. There are a number of configurations of the \erscn~rotor within the cage, as determined by X-ray crystallography and photoluminescence (PL) spectroscopy~\cite{olmstead00,tiwari07}. 

In this letter, we report the EPR signature of Er$^{3+}$ within \erscnC, making this the first endohedral ion known to be directly excitable both by EPR and optical techniques~\cite{shinohararev,jones06}. The EPR spectra indicate two primary configurations with considerable axial $g$ anisotropy. The temperature dependence of the linewidths are consistent with an Orbach mechanism via a higher electronic state of the Er$^{3+}$ ion, allowing us to assign the EPR peaks to configurations measured in PL. In addition, we report the photoisomerisation of the configuration of \erscn~encapsulated within a \ceighty~cage. We find that there are two primary configurations of the \erscn~unit, whose populations may be manipulated by optical excitation: an excess of one configuration, obtained by thermal annealing, is converted into an excess of the other under illumination. 

\erscnC~was purchased from Luna Innovations, and was further
purified using high performance liquid chromatography (HPLC). 
The sample, containing $\sim 10^{15}$ \erscnC~molecules dissolved in toluene, was freeze-pumped in liquid nitrogen
for five cycles to deoxygenate the solvent and was sealed under
vacuum in a quartz EPR tube. CW EPR experiments were performed using an X-band Bruker EMX Micro
EPR spectrometer equipped with a rectangular TE$_{102}$ resonator (with optical access)
and a low temperature helium-flow Oxford ESR900 cryostat. Measurements were performed in the temperature range 5 -- 30~K. A Nd:YAG laser was used to provide 532~nm illumination with a power of about 15~mW. Photoluminescence excitation (PLE) spectra at 5~K on frozen \cstwo~solution were acquired using an InGaAs array detector (Princeton Instruments) mounted on a Spex 1877B triple spectrometer. A laser diode, tunable between 1420 and 1520~nm, was used for the excitation at a power of 18~mW. 


Charge transfer across the molecule results in the nominal ionic configuration Er$^{3+}$(Sc$^{3+})_2$N$^{3-}$@C$_{80}^{6-}$. 
The Er$^{3+}$~ion is commonly used in doped glasses and crystals to generate optical amplification media~\cite{Eramp} and has been studied extensively using EPR in a variety of hosts~\cite{Hayes59,ErEPR,ErEPR2, ErEPR3,ErEPR4,ErEPR5}. 
It is known to fluoresce in the near-infrared regime due to the allowed optical transitions between the ground state $^{4}I_{15/2}$ and first excited state $^{4}I_{13/2}$. Photoluminescence spectroscopy reveals that the ground state of Er$^{3+}$~in this system is split into eight Kramers doublets, which can then be further split by an external magnetic field~\cite{abragam70,jones05}. Extensive mixing of $m_J$ within these states enables the observation of EPR between the doublets, which are traditionally treated as pseudo-spin 1/2 systems with a large effective $g$ factor~\cite{Hayes59}.

\begin{figure}[t] \centerline
{\includegraphics[width=3in]{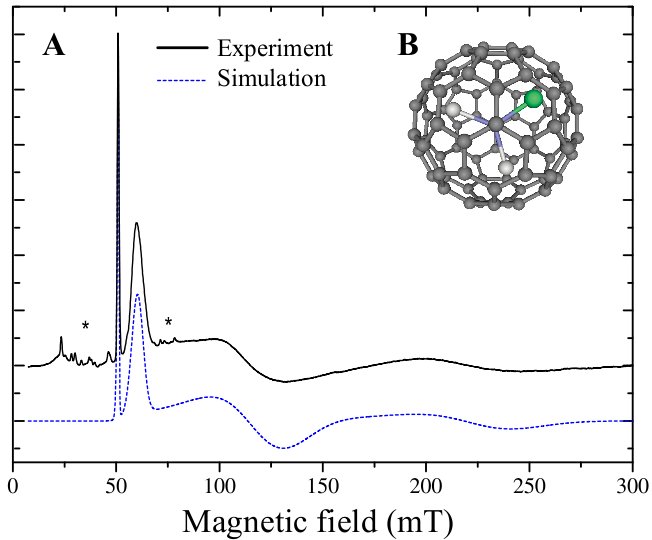}} \caption{(A) X-band CW EPR spectrum of \erscnC~in frozen
toluene solution at 5~K, with accompanying simulation performed using EasySpin~\cite{easyspin}. The low-intensity sharp features marked * derive mostly from hyperfine coupling to the $^{167}$Er nuclear spin (natural abundance 22.9 \%). Experimental parameters: microwave frequency, 9.396~GHz, microwave power, 63~mW; modulation amplitude, 0.3~mT; modulation frequency, 100~kHz. B) A representation of the \erscnC~molecule.}\label{fig1}
\end{figure}

\begin{figure}[t] \centerline
{\includegraphics[width=3.5in]{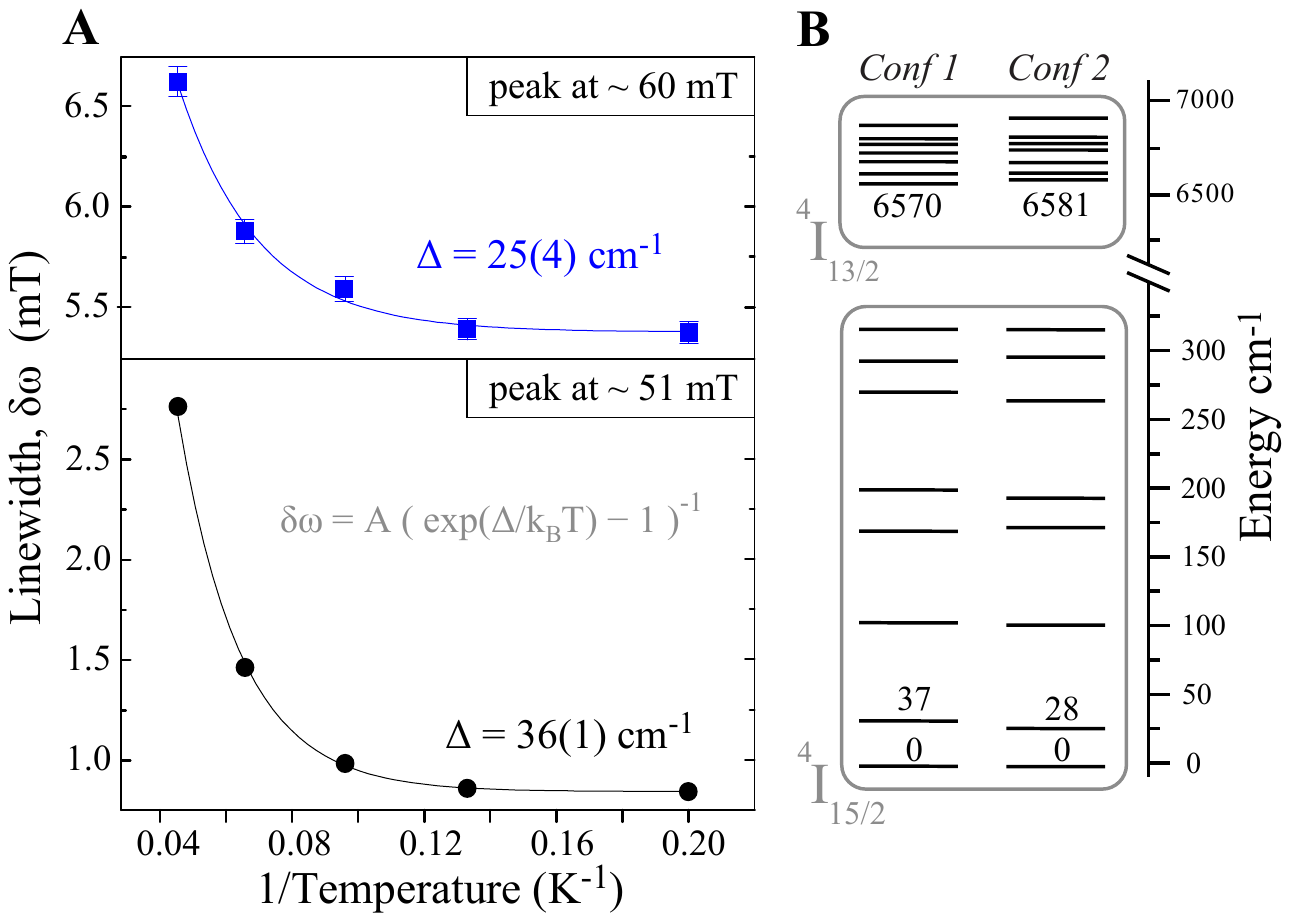}} \caption{(A) Linewidth of the two highest intensity features of the CW EPR spectrum shown in \Fig{fig1}, at approximately 51 and 60~mT. Their respective energy gaps $\Delta$ are obtained under the assumption that the features are homogeneously (lifetime) broadened by an Orbach mechanism (see \Eq{eq:orb}). These gaps match the energy level diagram (obtained from PL measurements) shown in (B) for the two major configurations of the \erscn~rotor inside the cage~\cite{tiwari07}.
}\label{fig:lw}
\end{figure}

The CW EPR spectrum of \erscnC~in frozen toluene solution at 5~K shown in \Fig{fig1}.
The widths of the two most intense features (at $\sim51$ and 60~mT) are plotted as a function of temperature in \Fig{fig:lw}. 
Assuming that the features are homogeneously broadened and that the transverse relaxation time \ttwo~is limited only by the spin-lattice relaxation time \tone~then the width $\delta\omega$ is proportional to \tone$^{-1}$. A common mechanism for \tone~relaxation in such systems with low-lying excited states is the two-phonon Orbach process~\cite{orbachEr}, for which
\beq
\delta\omega 
\propto T_1^{-1} 
\propto (e^{\Delta / k_B T}-1)^{-1}
+{\rm const.},
\label{eq:orb} 
\eeq
where $\Delta$ represents the energy of the excited state which forms the relaxation pathway. The close correlation between $\Delta$ (obtained from fits to Equation~\ref{eq:orb} shown as solid lines in \Fig{fig:lw}) and the energy of the first excited state as determined by PL~\cite{tiwari07} suggests that these two features in the EPR spectrum are associated with the two principal configurations of \erscn~within the fullerene cage. In order to explain the full EPR spectrum we similarly assign the broader features at higher field, though their weak intensity at 5~K makes the analogous temperature study more challenging.

\begin{figure}[t] \centerline
{\includegraphics[width=3in]{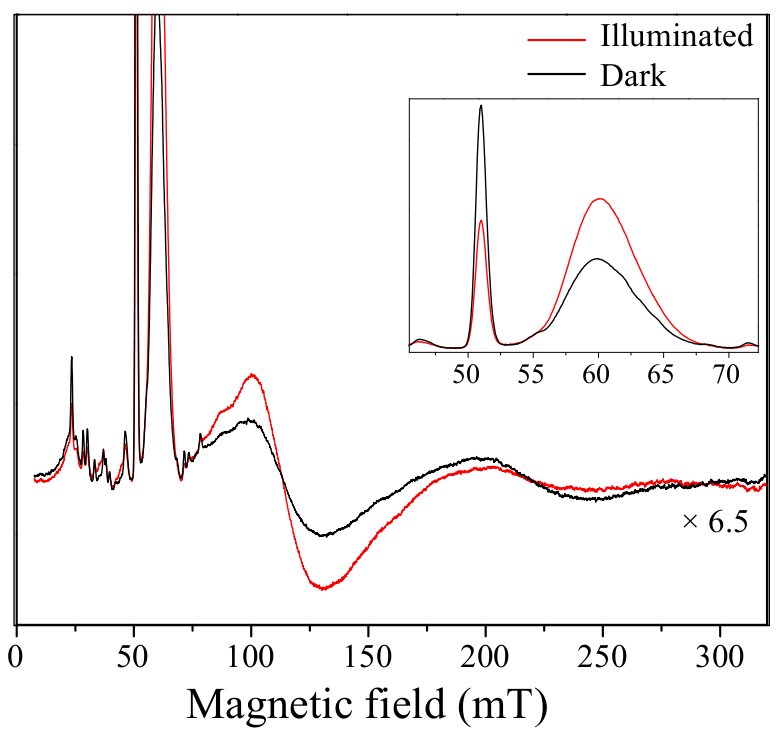}} \caption{CW EPR spectra taken at 5~K in the dark, and upon illumination with a 532~nm laser (inset expands the low-field region 45--75~mT). By pairing peaks according to whether their intensities increase or decrease upon illumination, we assign the features centred at approximately 51 and 220~mT to \sitea, and those at 60 and 110~mT to \siteb.}\label{fig:lightdark}
\end{figure}

Photoluminescence (PL) measurements demonstrate that the fullerene cage (and, as a result, the Er$^{3+}$ ion) can be excited at 532~nm. \Fig{fig:lightdark} shows the effect of such illumination on the CW EPR spectrum of \erscnC. The intensities of the principal features of the spectrum show a marked dependence on illumination: those centred at $\sim 51$ and 220~mT decrease in intensity, while those at about $\sim 60$ and 110~mT increase. This confirms that the high-field features have the same origin as the sharper low-field features, and indicates which pairs correspond to each of the primary configurations (which we term here \sitea~and \siteb). The CW spectrum in \Fig{fig1} should therefore be modelled with two independent contributions, each generating one low-field feature and one broader high-field feature; and the relative intensities of the contributions reflect the relative populations of \sitea~and \siteb~configurations (see below).

Differences in the crystal field and spin-orbit coupling for the Er$^{3+}$ ions in the two configurations produce two different effective spin systems. Although each is strictly a high spin system with a large crystal field splitting, only the ground state doublet is occupied at the temperatures studied here; the effective spin may be conveniently represented by an S=1/2 spin with an effective $g$ tensor. The dashed line in \Fig{fig1} shows a simulation of the powder spectrum of a two such spin systems: $g_{\perp}=3.0$,  $g_{\parallel}=13.67$ for \sitea, and $g_{\perp}=5.6$,  $g_{\parallel}=11.4$ for \siteb. The narrow features marked with an asterisk in \Fig{fig1} arise mostly from the hyperfine interactions with the 23\% abundance of $^{167}$Er ($I=7/2$), and some additional trace impurities in the resonator.  Since the state of the Er$^{3+}$ ion in this system may be measured by either photoluminesence or EPR, it may be possible to detect the magnetic resonance optically~\cite{odmrEr}.

This interpretation implies that the relative populations of the two configurations of \erscn~within the cage change under illumination at 532~nm. The change in spectra shown in \Fig{fig:lightdark} is metastable, and persists for many hours after the illumination ceases. The original `dark' spectrum can be recovered by annealing at $\sim 30$~K. Direct excitation of the Er$^{3+}$ ion at 1496 and 1499~nm has no effect on the EPR spectrum, suggesting that the configuration is affected only by excitation of the fullerene cage or the molecule as a whole, and not by excitation of individual Er$^{3+}$ ions. The magnitude of the effect shows a weak dependence on illumination wavelength in the region $<550$~nm (the magnitude of the effect at 550~nm is $\approx70\%$ that at 400~nm).

In order to deduce the change in populations between \sitea\ and \siteb\ from the change in the EPR spectrum under illumination, it is first necessary to relate the EPR intensity to the spin number. This relationship depends on many factors, including transition moment, temperature, microwave power and relaxation times. These terms can be combined into a constant of proportionality $\kappa$ such that $I_{i} =\kappa_{i} \, N_{i}$, where $I_i$ represents the intensity of any feature of the spectrum belonging to configuration $i$, and $N_i$ is the number of molecules in configuration $i$. Then, under the assumption that the total number of spin-active molecules does not change under illumination, the ratio $\kappa_1/\kappa_2$ may be extracted, and thus the fractional occupancy of the two sites before and after illumination. The results are summarised in Table~\ref{tab1}, which examines the dependence of the two sharp features at 51 and 60~mT as a function of illumination.

\begin{table}
{\includegraphics[width=3.5in]{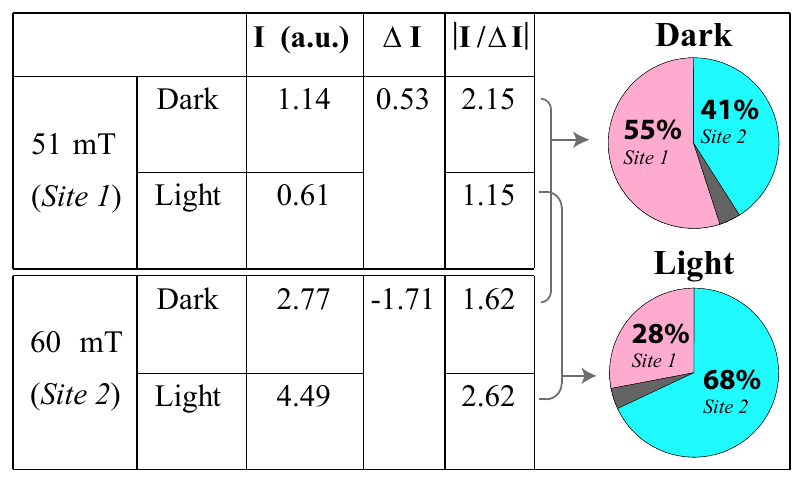}}
\caption{The integrated areas $I$ of the highest intensity EPR peaks at 51 and 60~mT are compared for spectra taken in the dark and under illumination at 532~nm. Assuming the number of spins contributing to the spectrum remains constant, the change in area $\Delta I$ can be used to normalise the peak intensity to produce a value proportional to spin number, and thus extract the occupancy of the two configurations of the \erscn~rotor within the fullerene cage. The small (4\%) region of each pie chart represents the estimated uncertainty resulting from experimental errors.}\label{tab1}
\end{table}

These results lead us to the remarkable conclusion that illumination turns a slight excess of \sitea~into a clear excess of \siteb. It appears the \sitea~is marginally favourable in the dark after annealing, but illumination at wavelengths of 532~nm or shorter switches the rotor preferentially to \siteb, where it remains metastable for many hours. 

Apparently, illumination at 532~nm generates an excited electronic state (either of the fullerene itself, or by driving charge transfer between the rotor and cage) in which \siteb\ is substantially favourable.
The origin of such a bias is an intriguing question, and may require detailed molecular modeling in order to yield a definitive answer. X-ray diffraction studies have been performed where the \erscnC~is chemically modified to co-crystallise into host~\cite{olmstead00}. That study revealed two principal configurations of the rotor present across the whole sample which were highly correlated with the two different fullerene orientations arising from different cage functionalisations. It is plausible that the different electronic distributions in these two different cage sites influence the dominant configuration of the internal species. This is consistent with our observation that for pristine cages optical excitation can lead to a switching of the majority species.  
It is known that the \erscn~rotor exhibits an electric dipole moment caused by the differences in electronegativity of erbium and scandium, as evidenced by a high degree of order observed in the XRD study of Ref~\cite{olmstead00}. 
Whether illumination creates an exciton on the fullerene or drives a charge transfer between the central species and the cage, the electric dipole associated with the cage may assist in producing the observed alignment.

\begin{figure}[t] \centerline
{\includegraphics[width=3in]{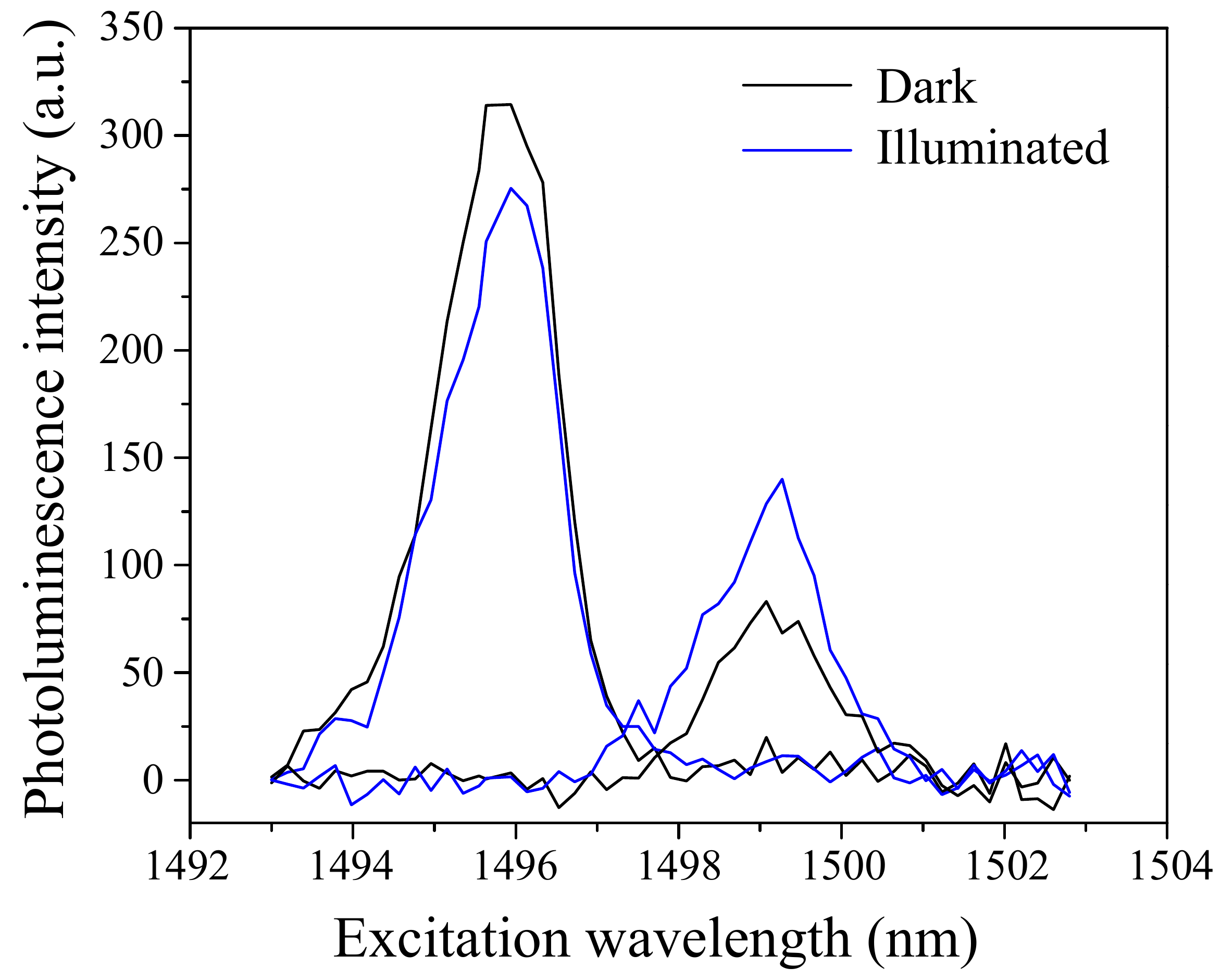}} \caption{PLE spectra of \erscnC~in frozen \cstwo~solution at 5~K, before (and after) having been illuminated at 532~nm for 20 minutes prior to data acquisition. The peak attributed to \sitea~(1496~nm) weakens upon illumination while that attributed to \siteb~(1499~nm) increases in intensity, consistent with what is observed in EPR (see \Fig{fig:lightdark}). 
}\label{fig:ple}
\end{figure}

The PLE experiments described in Ref~\cite{tiwari07} permit the selective excitation of Er$^{3+}$ in the two sites, and could therefore also be used to observe the photoisomerisation described above. Alignment of the optical apparatus was performed with a 532~nm laser beam, as is common practice, at 50~K (i.e.\ above the configuration annealing temperature of about 30~K found from EPR experiments above). The sample was then cooled down to 5~K and excited with an infra-red laser. For the two configurations, the PLE spectra were acquired while monitoring the fluorescence at respectively 1522~nm (PL peak attributed to \sitea) and 1519~nm (PL peak attributed to \siteb). In the range 1490--1500~nm, PLE peaks were found at 1496~nm for \sitea~and 1499~nm for \siteb.
532~nm illumination was then applied for 20 minutes at 5~K and turned off, and a second set of PLE spectra were taken. The results are shown in \Fig{fig:ple} and unambiguously demonstrate a a shift in site occupancy, as observed in EPR, though the noise level in this experiment precludes a quantitative comparison of occupancies.

In summary, the EPR spectrum of \erscnC~reveals two species of Er$^{3+}$, each with a high degree of apparent axial anisotropy (as expected for the case where the crystal field is much larger than the Zeeman splitting), corresponding to two configurations of the \erscn~rotor within the fullerene cage. The relative occupancies of these two orientations can be optically switched, as demonstrated in both EPR and PLE studies. The switching speed and reversibility of this photoisomerisation are yet to be determined; however, the long lifetime (over 12 hours at 20~K) of the switched orientation may find application in molecular memory elements.

We thank William Hayes and Alexei Tyryshkin for valuable discussions. This research is supported by the EPSRC through the QIP IRC www.qipirc.org (GR/S82176/01), and the Oxford Centre of Advanced Electron Spin Resonance (EP/D048559/1). JJLM is supported by St. John's College, Oxford. AA is supported by the Royal Society. GADB is supported by the EPSRC (GR/S15808/01).

\bibliography{bib}

\end{document}